\documentclass[aps,prl,twocolumn,superscriptaddress,amsfonts,amsmath,amssymb]{revtex4}
\usepackage{bm}
\usepackage{graphicx}
\usepackage{amsmath}
\usepackage{amstext}
\usepackage{amssymb}
\usepackage{amsfonts}
\usepackage{amsbsy}

\begin{document}

\title{Deconfined quantum criticality and Neel order via dimer disorder}
\author{Michael Levin}
\affiliation{Department of Physics, Massachusetts Institute of
Technology, Cambridge, Massachusetts 02139}
\author{T. Senthil}
\affiliation{Department of Physics, Massachusetts Institute of
Technology, Cambridge, Massachusetts 02139}
\begin{abstract}
Recent results on the nature of the quantum critical point between
Neel and valence bond solid(VBS) ordered phases of two dimensional
quantum magnets are examined by an attack from the VBS side. This
approach leads to an appealingly simple physical description, and
further insight into the properties of the transition.
\end{abstract}

\maketitle

Recent theoretical work\cite{dqcp-science,dqcp-longpaper} on quantum phase transitions in two
dimensional spin-$1/2$ quantum antiferromagnets has unearthed some
interesting phenomena dubbed `deconfined quantum criticality'. The
theory of such deconfined quantum critical points is described in
terms of excitations that carry fractionalized quantum numbers
which interact through an emergent gauge field. A precise
characterization of the deconfinement is provided by the emergence
of an extra global topological $U(1)$ symmetry not present at a
microscopic level. This symmetry leads to an extra conservation
law at the critical fixed point that is conveniently interpreted
as the conservation of a gauge flux.

The most prominent example of such a deconfined quantum critical
point arises at the transition between Neel and valence bond
solid(VBS) ordered phases of spin-$1/2$ magnets on a square
lattice. A direct second order transition is possible between
these two phases despite their very different broken symmetries, and in
contrast to naive expectations based on the Landau paradigm for
phase transitions. Previous results\cite{dqcp-science,dqcp-longpaper} on this transition have been
based primarily on an attack starting from the Neel ordered side.
Here we will take the alternate approach of attacking from the VBS
side. This new approach provides for an appealingly simple
physical description of the transition.

\begin{figure}[tb]
\centering
\includegraphics[width=3.0in]{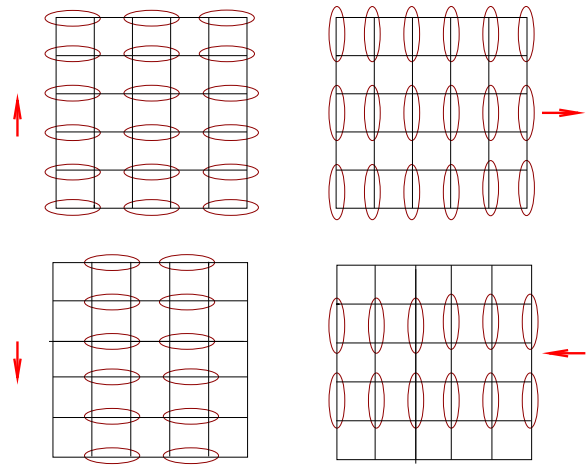}
\caption{Schematic picture of the four degenerate ground states
associated with the columnar VBS state. The encircled lines
represent the bonds across which the spins are paired into a
valence bond. The four ground states may be associated with four
different orientations of a $Z_4$ clock order parameter.}
\label{dimsol}
\end{figure}

The Neel ordered state is described by an $O(3)$ vector order
parameter. The VBS state, on the other hand, is described by a
$Z_4$ clock order parameter. The four degenerate ground states
associated with the $Z_4$ order parameter are illustrated in Fig. \ref{dimsol}
for a specific VBS state in which the valence bonds have lined up in columns. A naive approach to the transition from
the Neel side would associate the critical fixed point with the
usual $O(3)$ fixed point in $D = 2+1$ dimensions. This expectation
is incorrect. Similarly a naive approach to the transition from
the VBS side would lead one to expect a critical fixed point in the
$Z_4$ universality class in $D = 2 + 1$. This expectation is again
incorrect. As is well-known the {\em classical} $Z_4$ transition
in three dimensions is actually in the $D = 3$ $XY$ universality
class as the four-fold clock anisotropy is irrelevant at the
latter fixed point (for instance, see Ref. \cite{vicari}). The critical theory discussed in Ref.
\cite{dqcp-science,dqcp-longpaper} is emphatically not in the $3D$ $XY$ universality
class.

Why do these naive expectations fail? The answer is rooted in the
observation that the topological defects in either order parameter
carry non-trivial quantum numbers. When the defects in one order
parameter, say the Neel vector, proliferate and condense they kill
long range Neel order. At the same time the quantum numbers they
carry induces a different broken symmetry. This non-trivial
structure of the defects is inherently quantum mechanical and is
not captured in naive macroscopic treatments of the broken
symmetry state. For the Neel ordered states, the structure of the defects (known as hedgehogs)\cite{hald88}, and their role in producing
the VBS ordered paramagnet\cite{ReSaSuN} was elaborated many years ago. This provided the basis for the theory of the transition developed in Ref. \cite{dqcp-science,dqcp-longpaper}.
Here we will expose this physics starting from the VBS side.

\begin{figure}[tb]
\centering
\includegraphics[width=3.0in]{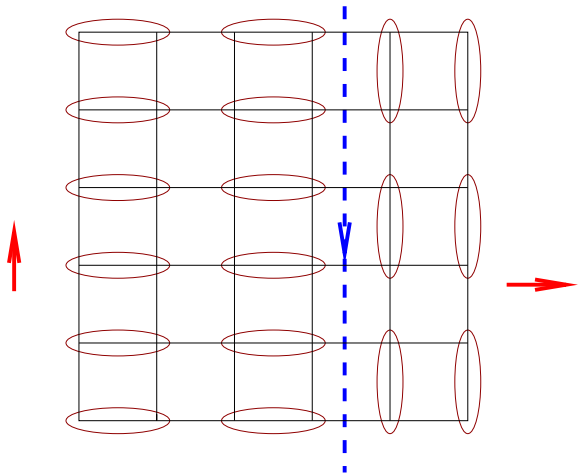}
\caption{An example of an elementary domain wall in the VBS state
across which the clock angle shifts by $\pi/2$.} \label{vbsdw}
\end{figure}

As the VBS order is described by a discrete $Z_4$ clock order
parameter, the natural topological defects are domain walls.
\footnote{Gauge theory aficionados will recognize that the domain
walls may be understood as electric field lines of the dual $Z_4$
gauge theory.} Various kinds of walls between the four different
broken symmetry states are possible. It is convenient to consider
an `elementary' domain wall across which the clock angle shifts by
$\pi/2$, and to assign an orientation to such a wall. An example
is shown in Fig. \ref{vbsdw}. All other walls, where the clock
angle shifts by higher multiples of $\pi/2$, may be built up from
the elementary wall.

\begin{figure}[tb]
\centering
\includegraphics[width=3.0in]{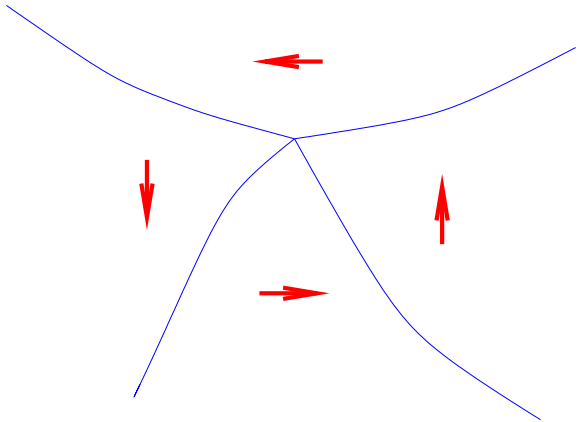}
\caption{Macroscopic picture of a $Z_4$ vortex as a point where four oriented elementary domain walls meet and end.} \label{z4vrtx}
\end{figure}

A key point is that $4$ such elementary walls can come together
and terminate at a point. In a macroscopic description focusing
only on the order parameter this is illustrated in Fig. \ref{z4vrtx}. It is
clear that such termination points may be associated with $Z_4$
vortices - the clock angle winds by $2\pi$ upon encircling such a
termination point. $Z_4$ antivortices may also be similarly defined.

\begin{figure}[tb]
\centering
\includegraphics[width=3.0in]{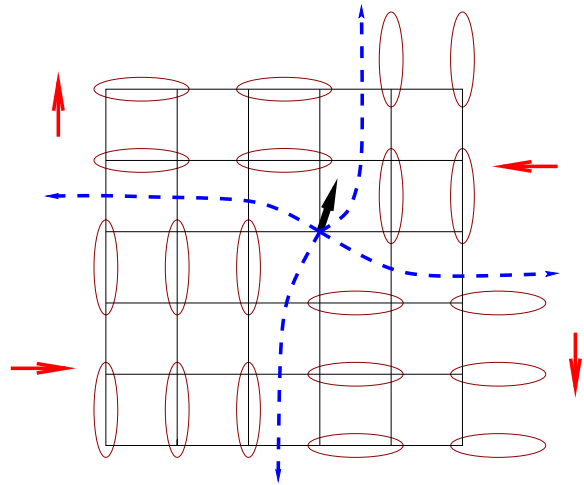}
\caption{The $Z_4$ vortex in the columnar VBS state. The blue lines represent the four elementary domain walls. At the core of the vortex there is
an unpaired site with a free spin-$1/2$ moment.} \label{vbsvrtx}
\end{figure}

What do such $Z_4$ vortices correspond to in terms of the
underlying VBS configurations? An example is illustrated in Fig. \ref{vbsvrtx}.
A remarkable property of this cartoon is that at the `core' of such
a vortex there is a site with an unpaired spin - {\em i.e} a spin
that is not part of any valence bond. It is easy to see that this
is a general property of any such vortex pattern of the VBS order
parameter. Furthermore, translating the entire valence bond pattern by one lattice spacing reverses the direction of the winding -
thus the $Z_4$ vortices are associated with
one sublattice, say the $A$ sublattice, and the $Z_4$ antivortices with the $B$ sublattice.

Thus in this particular quantum problem, the $Z_4$
vortices (and antivortices) carry an uncompensated spin-$1/2$ moment. They may
therefore be identified with `spinons'. In the VBS ordered phase the energy
cost of two such $Z_4$ vortices increases linearly with their
separation at long distances. Thus the spinons are {\em confined}
and do not exist as free excitations.

It is the non-trivial structure of the $Z_4$ vortex in this problem that
distinguishes the VBS state from a more ordinary state with a
$Z_4$ order parameter. Such an ordinary state obtains for instance
in a simple lattice quantum $O(2)$ rotor model with a four-fold
anisotropy. In this case the $Z_4$ vortices in the ordered state
have featureless cores. The disordering transition in this simple
model may be described by the usual three
dimensional classical $Z_4$ model and is hence in the $3D$ $XY$
universality class (since the clock anisotropy is irrelevant). In
contrast, disordering transitions out of the VBS phase must
necessarily take into account the presence of the spin-$1/2$
moment in the cores of the $Z_4$ vortices. Any mapping to a
classical $3D$ $Z_4$ model is then complicated by the need to
incorporate this vortex structure.

Consider moving out of the VBS phase by proliferating and
condensing the $Z_4$ vortices. Clearly once the vortices
proliferate long ranged $Z_4$ order cannot be sustained. Furthermore,
as these vortices carry spin, the resulting state will break spin
symmetry, and as argued below may be identified as the Neel state.

These simple considerations therefore provide a mechanism for a
direct second order transition between the VBS and Neel phases.
As for the usual $Z_4$ model, it
is reasonable to expect that  the
clock anisotropy will be irrelevant at this transition as well.
Indeed as we will argue later this is strongly supported by the
evidence from Ref.\cite{dqcp-science,dqcp-longpaper}. For the present let us explore the consequences
of the expected irrelevance of the clock anisotropy.

\begin{figure}[tb]
\centering
\includegraphics[width=3.0in]{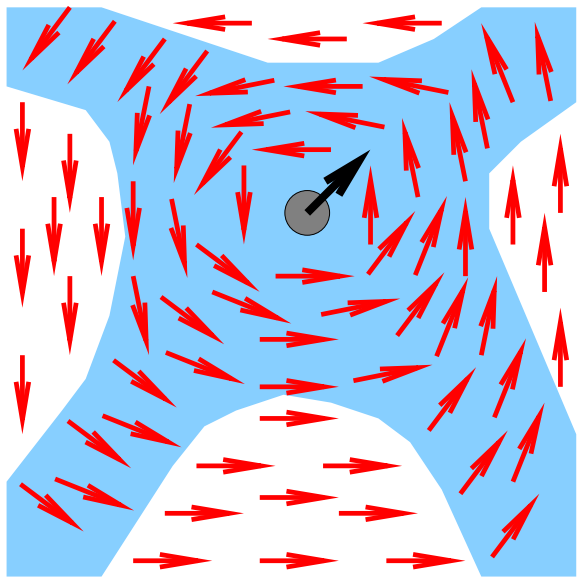}
\caption{The structure of the $Z_4$ vortex close to the transition. The domain walls have become thick (the blue shaded region). At length scales smaller than
the domain wall thickness $\xi_{VBS}$, the clock angle winds smoothly, as in
a regular $XY$ vortex. The core of the vortex, where the spin-$1/2$ moment resides, is characterized by a much smaller length scale \protect~$\xi$.
Both the domain wall thickness \protect~$\xi_{VBS}$ and the vortex core size \protect~$\xi$
diverge at the transition but the former diverges faster.} \label{z4xyvrtx}
\end{figure}

The critical theory will then be that of a (quantum) $XY$ model in
$D = 2+ 1$ but with vortices that carry spin-$1/2$ (See Fig. \ref{z4xyvrtx}).
The spinon nature of these vortices will change the universality class from
$D = 3 XY$ to something different. Clearly to expose this
difference and to obtain a description of the resulting new
universality class, it will be most convenient to go to a dual
basis in terms of the vortices and their interactions (analogous
to the familiar Coulomb gas description of {\em classical} $2D XY$
models).

The structure of such a dual vortex reformulation is well-known.
The basic idea is to regard the phase mode of the $XY$ model as
the `photon' associated with a fictitious non-compact $U(1)$ gauge
field. The vortices then correspond to gauge charges that are
minimally coupled to this photon field. At the critical point, the
vortices are gapless: the critical theory may be constructed as a
theory of gapless vortex fields minimally coupled to a fluctuating
non-compact $U(1)$ gauge field.
For the problem at hand, the spinon nature of the vortices is
readily incorporated by introducing a two-component spinor field
$z_a$ to represent the vortices ($a =1,2$ is the spin index). The transition out of the VBS phase to the Neel phase
will then be described by a theory of gapless spinon-vortices coupled minimally to a fluctuating non-compact $U(1)$ gauge field (which is the dual of the
$XY$ phase mode).

The critical theory is readily written down. The most general theory
consistent with the $U(1)$ gauge structure, $SU(2)$ symmetry, and
vortex/antivortex exchange symmetry of the microscopic model (the latter
required by the symmetry under sublattice exchange $A \leftrightarrow B$),
is described by the action $\mathcal{S}_z = \int
d^2 r d \tau \mathcal{L}_z$, with
\begin{eqnarray}
\mathcal{L}_{z} &=&  \sum_{a = 1}^2 |\left(\partial_{\mu} -
ia_{\mu}\right) z_{a}|^2 + s |z|^2 +
u\left(|z|^2 \right)^2 \nonumber \\
&~&~~~~~~~ +
\kappa\left(\epsilon_{\mu\nu\kappa}\partial_{\nu}a_{\kappa}\right)^2,
\label{sz}
\end{eqnarray}
The transition occurs as the parameter $s$ is tuned. The $a_{\mu}$
represent the components of a fluctuating gauge field.

Remarkably this is exactly the same field theory as the one proposed in Ref. \cite{dqcp-science,dqcp-longpaper}
for the Neel-VBS transition based on an approach that attacked
from the Neel side. We have thus shown how to recover that field
theory in an approach from the VBS side.

These considerations may be formalized as follows. First we note that $z_a$ represents a $Z_4$ spinon-vortex, and hence must transform as a spinor
under physical SU(2) spin rotations. The antivortex must also transform as a spinor - we must therefore represent antivortices by
$-i\sigma^y_{ab}z_b^*$ where $\sigma^y$ is the usual Pauli matrix. As discussed pictorially above, elementary lattice translations take vortices to antivortices so that
$z_a \rightarrow -i\sigma^y_{ab}z_b^*$. It follows that the vector $\vec N = z^{*}\vec \sigma_{ab} z_b$ changes sign under an elementary lattice translation.
We may therefore identify it with the Neel order parameter. Thus for instance a uniform condensate of $z_a$ corresponds to the Neel state.

We may formally justify the critical theory in Eqn.\ref{sz} above as follows. The arguments developed above show that the critical theory is
that of an $XY$ model where the vortices are spinons. Consider the conserved current $J_{\mu}$ of this $XY$ model. In the ordered
phase this may be expressed in terms of the
$XY$ phase field $\chi$ through
\begin{equation}
J_{\mu} = K\partial_{\mu} \chi
\end{equation}
where $K$ is the stiffness of the $XY$ model. To access the $XY$ disordered phase, it is necessary to include vortex configurations
and account for the periodicity of the phase $\chi$. The vortex current $j_{\mu}$ is given by
\begin{equation}
j_{\mu} = \frac{1}{2\pi} \epsilon_{\mu \nu \lambda}\partial_{\nu}\partial_{\lambda}\chi
\end{equation}
Note that $j_{\mu}$ must be invariant under physical spin rotations even though it is carried by spinons.
The conservation condition on $J_{\mu}$ may be implemented by expressing it as
\begin{equation}
J_{\mu} = \frac{1}{2\pi}\epsilon_{\mu \nu \lambda}\partial_{\nu}a_{\lambda}
\end{equation}
This equation defines the field $a_{\mu}$ which may be interpreted as a non-compact $U(1)$ gauge field. Clearly it is defined only up to a
gauge transformation $a_{\mu} \rightarrow a_{\mu} + \partial_{\mu}\theta$. The vortex current may now be re-expressed in terms of $a_{\mu}$:
\begin{equation}
j_{\mu} = \frac{1}{4 \pi^2 K}\epsilon_{\mu \nu \lambda}\partial_{\nu}b_{\lambda}
\end{equation}
where $b_{\lambda} = \epsilon_{\lambda \alpha \beta}\partial_{\alpha}a_{\beta}$ is the gauge-invariant field strength associated with the $a_{\mu}$
field. This equation now takes the form of the familiar Ampere law. The duality is completed by requiring a continuum field theory of the spinon-vortices $z_a$
whose equations of motion reduce to this Ampere law equation. The action in Eqn. \ref{sz} above has precisely this property as is readily checked.

Note that as usual the conserved density $J_0$ of the $XY$ model is simply the magnetic flux density in the dual description.
As the phase $\chi$ is the conjugate operator, the operator $e^{4i\chi}$ simply increases the total gauge flux by $8\pi$.
We may therefore identify it with a quadrupled monopole operator of the dual gauge theory. Thus
the quartic anisotropy in the $XY$ model corresponds precisely to the quadrupled monopole operator. Strong evidence for the irrelevancy of this operator
at the critical fixed point of Eqn. \ref{sz} was presented in Ref. \cite{dqcp-longpaper}.

Is it possible for quantum fluctuations to destroy the VBS order
without inducing Neel order? Clearly the answer is yes. One
possibility is a transition to a topologically ordered $Z_2$ spin
liquid state. To obtain a topologically ordered state from the VBS
state it is as usual necessary to condense {\em paired}
vortices\cite{BFN} in the VBS order parameter - but here these
vortices are spinons. To get a spin singlet state it is necessary
to form a singlet pair of these spinons. \footnote{Readers
familiar with the gauge theoretic description of the $Z_2$ spin
liquid will recognize that this is the same as the usual procedure
of condensing a gauge charge-$2$ spinon pair to obtain the $Z_2$
phase.} As discussed above, vortices live on one sublattice and
antivortices on another. Consequently we need to condense a
spin-singlet pair of spinons living on the same sublattice to
obtain the $Z_2$ spin liquid. All of this is completely consistent
with existing gauge theoretic descriptions of $Z_2$ spin
liquids\cite{RSSpN}.

To conclude, we have examined the nature of the Neel-VBS transition by an attack from the VBS side.
This approach leads to a simple
physical description of the transition and is completely consistent with the alternate approach of
attacking from the Neel side.
All of the physics associated with the transitions out of the VBS phase may be fruitfully
understood from the perspective of this paper.

\begin{acknowledgments}
This research is supported by the National Science Foundation
grant DMR-0308945 (T.S.).  T.S also acknowledges funding from the
NEC Corporation, the Alfred P. Sloan Foundation, and an award from
The Research Corporation.

\end{acknowledgments}

\vspace{0.5in}

\bibliography{avbs}

\end{document}